\documentclass[reprint,amsmath,amssymb,aps]{revtex4-2}
\usepackage{graphicx}
\usepackage{dcolumn}
\usepackage{bm}

\begin{document}

\title{Stable electro-optic modulators using thin-film lithium tantalate}

\author{Keith Powell}
\email{kpowell@fas.harvard.edu}
\affiliation{John A. Paulson School of Engineering and Applied Sciences, Harvard University, Cambridge, Massachusetts, 02138, USA}

\author{Xudong Li}
\affiliation{John A. Paulson School of Engineering and Applied Sciences, Harvard University, Cambridge, Massachusetts, 02138, USA}

\author{Daniel Assumpcao}
\affiliation{John A. Paulson School of Engineering and Applied Sciences, Harvard University, Cambridge, Massachusetts, 02138, USA}

\author{Let\'{i}cia Magalh\~{a}es}
\affiliation{John A. Paulson School of Engineering and Applied Sciences, Harvard University, Cambridge, Massachusetts, 02138, USA}

\author{Neil Sinclair}
\affiliation{John A. Paulson School of Engineering and Applied Sciences, Harvard University, Cambridge, Massachusetts, 02138, USA}

\author{Marko Lon\v{c}ar}
\email{loncar@seas.harvard.edu} 
\affiliation{John A. Paulson School of Engineering and Applied Sciences, Harvard University, Cambridge, Massachusetts, 02138, USA}

\begin{abstract}
We demonstrate electro-optic modulators realized in low-loss thin-film lithium tantalate with superior DC-stability ($<1$ dB power fluctuation from quadrature with 12.1 dBm input) compared to equivalent thin-film lithium niobate modulators (5 dB fluctuation) over 46 hours.
\end{abstract}

\maketitle
\noindent

\section{Introduction}
Thin-film lithium niobate (TFLN) integrated electro-optic (EO) circuits have bolstered many advancements in optical science and technology in the last several years \cite{zhu2021}.
The most seminal EO device--the Mach-Zehnder modulator (MZM)-- benefits a host of technologies in optical communications, sensing and computing.
State-of-the-art figures-of-merit for TFLN MZMs at 1550 nm include half-wave voltage length product (V$_{\pi}$L) and optical loss of 2.3 Vcm and 3 dB/m, respectively, and EO bandwidths beyond 100 GHz \cite{zhu2021}. 
An unresolved issue with TFLN modulators is slow ($<$kHz) relaxation of the EO response. 
As a result, active EO control of the modulator is employed, or alternatively thermo-optic based control and biasing is utilized \cite{xu2020high,holzgrafe_TFLN_relaxation}.
The former, however, introduces additional complexities, while the latter results in increased steady-state power consumption and is not compatible with cryogenic operation. 
Furthermore, at elevated optical powers the photorefractive effect can introduce additional relaxation dynamics and instabilities. 
\cite{zhu2021}.

Lithium tantalate (LT) is known to have slower EO relaxation than LN, while featuring comparable, or better, properties \cite{LT_damage,shen23,yu_24,wang2023}:
30~pm/V EO coefficient, 3.93 eV bandgap, much lower birefringence, photorefraction, and RF loss, as well as higher optical damage threshold.
Recently, the first optical devices on TFLT have appeared: microdisks \cite{LT_damage}, strip-loaded waveguides \cite{shen23}, as well as dry-etched microrings \cite{yu_24} and MZMs \cite{wang2023}.
Nevertheless, aside from Kerr combs leveraging the low birefringence \cite{wang2023} and recent attempts at measuring photorefraction \cite{yu_24}, none of these works show reason to use TFLT over TFLN for electro-optics.

We design, fabricate, and demonstrate TFLT EO MZMs operating at 1550 nm with superior DC stability compared to equivalent TFLN MZMs.
Importantly, this is achieved without sacrificing other device figures of merit including optical loss, RF bandwidth, and drive voltage.
We also measure racetrack resonators to further confirm optical loss is not compromised using our fabrication process.

\section{Fabrication}
TFLT devices are fabricated on 500 nm-thick x-cut TFLT-on-insulator (NanoLN). 
Waveguides are patterned using maN-2405 resist and 150 keV EBL, then etched 300 nm using Ar$^{+}$ ICP.
Redeposition is removed using a high-pH clean.
The devices are annealed in an O$_2$ atmosphere at 520°C for 2 h. 
The MZMs for DC-drift measurements are uncladded for comparing LT to LN.
For the other MZM measurements, an 800 nm-thick SiO$_2$  cladding layer is PECVD-deposited to improve RF-optical overlap. 
Cladding is removed by patterning with SPR700-1.0 resist, photolithography, and etching to LT using C$_3$F$_8$, O$_2$, and Ar$^{+}$ ICP.
E-beam evaporation and lift off define the electrodes (15 nm-thick Ti, 800 nm-thick Au) and termination (20 nm-thick Ti, 5 nm-thick Pt).
MZM devices are finally hotplate-heated at 300°C for 5 h.
To unambiguously compare the TFLT MZM with TFLN, we fabricate MZMs in TFLN with the same geometry.

\begin{figure*}
\includegraphics[width=\linewidth]{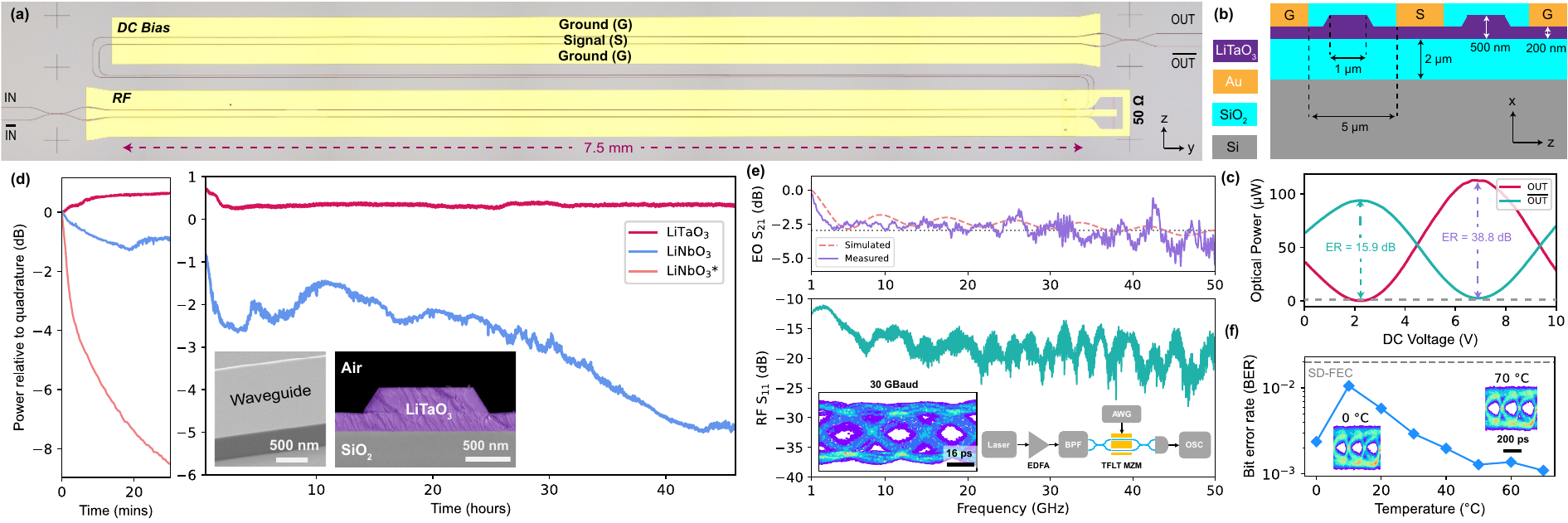}
\caption{Integrated thin-film lithium tantalate-on-insulator electro-optics. (a) Micrograph of a balanced Mach-Zehnder modulator with independent DC and 50 Ohm-terminated RF electrodes. (b) Modulator cross-section schematic with crystal axes and relevant dimensions. (c) Modulator transfer function at each output with slowly varied applied voltage. (d) Dynamics of relative output intensity when bias set to quadrature and comparison with thin-film lithium niobate modulators over 46 hours. First 30 min is shown in left plot. Curve (*) is from Ref. \cite{xu2020high}. 
Insets: electron micrograph of the waveguide (left) and false-colored electron micrograph of the cross-section of the waveguide with 60° sidewall angle (right) . 
(e) Normalized electro-optic conversion efficiency S$_{21}$, measured (simulated) shown using a solid (dashed) line, and RF S$_{11}$ spectra. Insets: Eye diagram for 30 GBaud non-return-to-zero pseudorandom bit sequence modulation using high optical power (left) and associated setup (right). EDFA: erbium-doped fiber amplifier, BPF: bandpass filter, AWG: arbitrary waveform generator, PD: photodector, OSC: oscilloscope. 
(f) Measured bit error rate is below soft-division forward error correction threshold under varied temperature. Insets: Associated eye diagrams at extrema. 
}
\label{fig:figure1}
\end{figure*}

\section{Results}
An optical image of the TFLT MZM is shown in Fig. \ref{fig:figure1}a.
We use a 2x2 path length-balanced MZM to mitigate temperature and laser wavelength fluctuations. 
Separate 7.5 mm-long GSG DC and 50 $\Omega$-terminated RF electrodes avoids an external bias-tee, thereby providing excellent isolation between DC and RF sources while enabling low RF insertion loss, flat RF response, and avoids terminator damage under high DC bias.
The MZM cross-section (Fig. \ref{fig:figure1}b schematic) is optimized for low optical loss with strong optical and RF field overlap.
Light is coupled to and from the MZMs using fibers and grating couplers with 7.9 dB loss each. 
We measure low V$_{\pi}$L of 3.4 Vcm at near-DC (Fig. \ref{fig:figure1}c), limited only by our current device geometry, including gap width, etch depth, etc, which can be optimized in future work.
Notably, an exceptionally high extinction ratio of 39 dB is measured.
The second port has a lower extinction ratio of 16 dB due to asymmetric directional couplers (53:47) and, as observed using a microscope, loss induced by misalignment of the electrode (0.7 dB).
The ratio and electrode-induced loss are estimated using simulation.
Scattering in the MZM waveguides accounts for 0.13 dB loss leading to a total on-chip loss of the MZM of 0.35 dB.
This is consistent with that inferred from a comparison measurement of a simple waveguide.

Next, the MZM is biased at quadrature and EO relaxation is measured over 46 hours using 20 dBm off- (12.1 dBm on-) chip input optical power (Fig. \ref{fig:figure1}d). 
We find $<$1 dB of laser intensity variation in the first two hours, likely due to charge migration at the electrical contact, settling to 0.2 dB drift over the final 44 hours.
On the other hand, the TFLN MZM exhibited 5 dB laser intensity variation over the entire timeframe when measured under the same conditions. 
Further, EO relaxation of TFLN reported in literature varies from 3 dB in $\sim$15 minutes \cite{holzgrafe_TFLN_relaxation} to 8 dB over $\sim$30 mins \cite{xu2020high}, optical powers unstated in both works.
Matching of the optical group and RF phase velocities produces a normalized MZM electro-optic conversion efficiency (Fig. \ref{fig:figure1}e) with a flat roll-off to 50 GHz after 3 GHz (limited by our network analyzer). 
The roll-off before 3 GHz is due the low characteristic impedance of the transmission line electrode (39 $\Omega$), which can be straightforwardly overcome by an optimized device geometry using a thicker bottom oxide.
The RF back-reflection is due to the aforementioned RF impedance mismatch.

To show benefit of our stable DC bias in conjunction with our path length-balanced design, we bias the MZM at quadrature and drive it using a non-return-to-zero (NRZ) pseudorandom binary sequence (PRBS) at 5 GBaud, over a 70°C temperature range, (Fig. \ref{fig:figure1}f). 
Eye diagrams are measured with 5.1 dBm on-chip optical power.
They yield bit error rates (BERs) less than that required for soft-division forward error correction (SD-FEC), corresponding to signal-to-noise ratios of $\sim$3.
The reduced error rate with increased temperature is due to thermal expansion-induced alignment improvement of the optical fiber with the grating couplers.
The increased error at low temperatures owes to misalignment induced by condensation. 
Errors can be kept low with packaging.
Furthermore at room temperature, and with high 28 dBm on-chip optical power, we retain open eyes at up to 30 GBaud modulation rates and a signal-to-noise ratio of 3.3, limited by our oscilloscope bandwidth (Fig. \ref{fig:figure1}e, inset).
Instabilities of the optical amplifier prevented accurately testing the long-term DC stability at this power.

To determine optical waveguide loss, we measure an uncladded racetrack resonator with a free spectral range of 185 pm.
We find an intrinsic Q factor of 4.04 million, corresponding to a FWHM linewidth of 47.2 MHz, and 9 dB/m waveguide propagation loss (including bending loss). 
This is comparable to the lowest loss achieved in TFLT, but fabricated using hard-mask transfer \cite{wang2023}, and within the same order of magnitude achieved in LN (3 dB/m).
To verify the film quality, we also measure a thermo-optic coefficient of $d{n}/dT = 2.47\times 10^{-5} K^{-1}$, which indeed matches the literature value of bulk LT \cite{Iwasaki_1968}, given our racetrack geometry.
Refractive indices $n_e \sim n_o \sim 2.12$ at 1550 nm are independently determined by ellipsometry.

\section{Conclusion} 
We present high-extinction TFLT MZMs that have improved DC stability over equivalent MZMs in TFLN, without sacrificing key properties including RF bandwidth, drive voltage, and optical loss.
Charge dynamics near surfaces play a significant role in the performance of nanoscale devices, and we expect that optimization of these effects could improve TFLT electro-optics beyond what is presented here.

\subsection*{Funding}
NSF EEC-1941583, AFOSR FA9550-20-1-01015,NSF 2138068,
NASA 80NSSC22K0262, MagiQ Technology/Naval Air Warfare Center N6833522C0413, Amazon Web Services.

\subsection*{Acknowledgments} The authors thank D. Renaud, D. Zhu, M. Yeh, D. Barton, and C.J. Xin.
This work was performed in part at the Harvard University Center for Nanoscale Systems (CNS); a member of the National Nanotechnology Coordinated Infrastructure Network (NNCI), which is supported by the National Science Foundation under NSF award no. ECCS-2025158.

\bibliography{main}

\end{document}